\documentstyle[prb,aps,epsf,array]{revtex}
\begin{document}
\draft

\title{Force, charge, and conductance of an ideal metallic nanowire}

\author{F.\ Kassubek} 
\address{Fakult\"at f\"ur Physik, Albert-Ludwigs-Universit\"at,
Hermann--Herder--Str.~3, D-79104 Freiburg, Germany}
\author{C.\ A.\ Stafford}
\address{Fakult\"at f\"ur Physik, Albert-Ludwigs-Universit\"at,
Hermann--Herder--Str.~3, D-79104 Freiburg, Germany}
\address{and}
\address{Physics Department, University of Arizona, Tucson, AZ 85721$^*$}
\author{Hermann Grabert}
\address{Fakult\"at f\"ur Physik, Albert-Ludwigs-Universit\"at,
Hermann--Herder--Str.~3, D-79104 Freiburg, Germany}

\date{\today} \maketitle

\begin{abstract}
The conducting and mechanical properties of a metallic nanowire formed
at the junction between two macroscopic metallic electrodes are investigated.
Both two- and three-dimensional wires with a W(ide)--N(arrow)--W(ide)
geometry are modelled in the free-electron 
approximation with hard-wall boundary conditions.   Tunneling and 
quantum-size effects are treated exactly using the scattering matrix 
formalism.  Oscillations of order $E_F/\lambda_F$ in the tensile force are 
found when the wire is stretched to the breaking point, which are synchronized
with quantized jumps in the conductance.
The force and conductance are shown to be essentially independent of the 
width of the wide sections (electrodes).  The exact results are compared with
an adiabatic approximation; the later is found to overestimate the effects of
tunneling, but still gives qualitatively reasonable
results for nanowires of length $L\gg\lambda_F$, even for this abrupt geometry. 
In addition to the force and conductance, the net charge of the nanowire
is calculated and the effects of screening are included within linear
response theory.  Mesoscopic charge fluctuations of order $e$ are predicted
which are strongly correlated with the mesoscopic force fluctuations.
The local density of states at the Fermi energy exhibits nontrivial
behavior 
which is correlated with fine structure in the force and 
conductance, showing the importance of treating the whole wire as a
mesoscopic system rather than treating only the narrow part.

\end{abstract}

\section{Introduction}
\label{sec.intro}

Metallic nanowires may be formed at the junction between two metallic 
electrodes which are pressed together and/or pulled apart in a controlled
fashion.\cite{nanowires,olesen,krans,nanoforce,nanoforce2,histogram}  
In a wire of nanoscopic cross section, the transverse motion is quantized,
leading to a finite number of electronic modes below the Fermi energy
$E_F$ which can be transmitted through the wire.  A striking consequence
of these discrete modes is the quantization of the wire's conductance 
at integer multiples of $G_0=2e^2/h$, a phenomenon 
first observed in two-dimensional (2D) semiconductor
heterostructures,\cite{vanwees,wharam,beenakker} 
and subsequently studied
in detail in three-dimensional (3D) metallic 
nanowires.\cite{nanowires,olesen,krans,histogram}
A subtlety inherent in conductance quantization experiments 
is that even for a nearly ideal nanowire, the
presence of disorder in the electrodes far from the region of
interest leads to a 
suppression of the conductance plateaus below integer values.  This 
suppression is normally taken into account by subtracting a phenomenological
series resistance,\cite{krans,histogram,vanwees,wharam} 
which allows one to shift
the experimentally observed plateaus back to integer values.  
Theoretical histograms \cite{bascones,hist-sim,jerome} exhibit a similar shift
towards lower values of the conductance, though the precise form of the
suppression is not equivalent to a simple series resistance
due to quantum interference effects. \cite{been,shesds}  These considerations
underline the importance of treating the nanowire and the electrodes connected
to it as a single mesoscopic
system, rather than considering the nanowire in isolation.

The cohesive properties of good metals are also determined to a large 
extent by the conduction electrons.  Thus, one may expect the mode quantization
in a nanowire to have a strong effect on its mechanical properties as 
well.\cite{sbb}
In a pioneering experiment published in 1996, 
Rubio, Agra\"{\i}t, and Vieira measured
simultaneously the force and conductance during the formation and rupture
of an atomic-scale Au nanowire.\cite{nanoforce}  They observed oscillations
in the tensile force of order 1nN under deformation, which were synchronized
with 
jumps of order $2e^2/h$ in the conductance.  Similar experimental results
were obtained independently by Stalder and D\"urig.\cite{nanoforce2}
In a previous paper,\cite{sbb} it was shown that this intriguing behavior
can be understood quantitatively using a simple free-electron jellium 
model for a metallic nanowire.
The theoretical approach introduced by Stafford, Baeriswyl, and
B\"urki\cite{sbb} uses the 
electronic scattering matrix to describe the coupling of the nanowire
of interest to the macroscopic probes ({\em e.g.}, STM/AFM)
used to manipulate it;  
the correct treatment of this coupling is crucial for calculating
the mesoscopic corrections to the bulk electrical and mechanical properties.
In the previous paper,
the scattering matrix was evaluated using the 
adiabatic and WKB approximations, appropriate for a smooth geometry in which
the diameter $D$ of the nanowire varies slowly along its symmetry axis $z$,
{\em i.e.}, $(d D/d z)^2 \ll 1$.
The qualitative picture which emerged from the analysis of Ref.\
\onlinecite{sbb} is that each quantized mode contributing $2e^2/h$ to the 
conductance of the nanowire also contributes an amount of order
$E_F/\lambda_F$ to its cohesive force, where $\lambda_F$ is the De Broglie
wavelength of electrons at the Fermi energy $E_F$.  
(For monovalent metals, $E_F/\lambda_F$ is of order 1nN.)
Under elongation, the cross section of the nanowire narrows, and each time a 
transverse mode is cut off, both the conductance and the cohesive force
decrease abruptly.
Using an elegant
argument based on the technique of Ref.\ \onlinecite{sbb}, 
H\"oppler and Zwerger showed that the leading order mesoscopic correction to
the tensile force depends only on the topology of the cross section of the 
nanowire.\cite{commentzwerger}
The scattering matrix formalism
has also been used to study the effects of impurity scattering on the
mechanical properties of nanowires.\cite{impurities}  

Independently,
van Ruitenbeek, Devoret, Esteve, and Urbina \cite{vRuitenbeek}
considered wires with a special W(ide)--N(arrow)--W(ide) geometry,
with two wide outer sections, representing the electrodes,
and a narrow inner section of constant diameter, representing the nanowire
(see Fig.\ \ref{fig.1}).  They considered the limit (cf.\ 
also Refs.\ \onlinecite{landman2,blom}) 
where the narrow section
is sufficiently long that the boundary effects at the junctions of the 
wide and narrow sections give a negligible contribution to the energetics
of the nanowire.
They also pointed out that screening should be included, and imposed a
charge neutrality constraint as a first approximation.
However, they noted that the assumption of local
charge neutrality of the nanowire breaks down for very short wires, such as
those investigated in the experiments of Refs.\ 
\onlinecite{olesen,krans,nanoforce,nanoforce2,histogram}, which are only
of order 1nm in length.
For such short wires, boundary effects are important.

In the present article, we investigate both 2D and 3D wires with the WNW
geometry, treating the boundary effects
arising from the connection of the nanowire to the electrodes exactly
via the scattering matrix approach. 
Our results for the force in the WNW
geometry are qualitatively similar to the results for smooth geometries
presented in Ref.\ \onlinecite{sbb}, although the presence of sharp corners
is shown to lead to a non-generic topological correction to the force for
small deformations from an ideal wire, which is well understood.
The exact results for the WNW geometry are compared with an
adiabatic approximation, which is found to overestimate the effects of
tunneling, but still gives qualitatively reasonable results for the conductance
and force of nanowires of length $L \gg \lambda_F$, even for this abrupt
geometry.  The local density of states at the Fermi energy is also
calculated, and is shown to exhibit strong modulation on a length scale
of order $\lambda_F$. This electronic structure of the scattering states
gives vivid pictorial support of the notion\cite{sbb}
that conductance channels act as metallic bonds.

In addition to the force and conductance, the net charge of the nanowire
is calculated, and the effects of screening 
are adressed by a linear response approach. 
We predict mesoscopic charge fluctuations of
order the 
fundamental charge quantum $e$,
which are synchronized with the quantized steps in
the wire's conductance, and strongly correlated with the mesoscopic
force fluctuations.  Similar charge fluctuations are predicted in 2D
and 3D nanowires; they should thus be present in quasi-two-dimensional 
quantum point contacts exhibiting conductance 
quantization\cite{vanwees,wharam,beenakker} as well.
The smallness of the predicted mesoscopic charge imbalance leads us to neglect
electron-electron interactions in our treatment of the conducting and 
mechanical properties of the nanowire.
Indeed, we find that the mesoscopic corrections to the force for wires 
of length $L \sim \lambda_F$ are large compared to the corrections expected
due to screening effects, justifying our approach.  

The paper is organized as follows: In Sec.\ \ref{sec.smatrix}, we review the 
scattering matrix formulation of electrical conduction and statistical
mechanics.  General expressions for the force, charge and conductance of 
a mesoscopic conductor in terms of the electronic scattering matrix are
derived.
In Sec.\ \ref{sec.comp_smatr}, the scattering matrix for 2D and 3D wires with 
the WNW geometry is calculated.  For simplicity, 3D wires with 
a square cross section are considered, but it is
straightforward to use our method for wires of arbitrary cross section if
the eigenfunctions and eigenvalues of the 2D Schr\"odinger equation with
Dirichlet boundary conditions for this shape are known.
In Sec.\ \ref{sec.2Dnanowire}, the results for the force, charge,
and conductance of
a 2D nanowire are presented.  The semiclassical approximation to the
force and charge, and the topological contribution due to sharp
corners are discussed.  The exact results are compared to an adiabatic
approximation, and the local density of states at the Fermi energy is 
calculated.
In Sec.\ \ref{sec.3dnanowire}, the results for the force, charge,
and conductance of 3D nanowires are presented.  The effects of screening
are evaluated within linear response.
The relevance of our results to the 
experiments of Refs.\ \onlinecite{nanoforce} and \onlinecite{nanoforce2} 
is discussed.
Some general conclusions are presented in Sec.\ \ref{sec.concl},
and a comparison of the jellium model used here to atomistic descriptions of
nanowires based on classical molecular dynamics simulations
\cite{landman,todorov,mol.dyn} is given.

\section{S-matrix formalism}
\label{sec.smatrix}

A nanowire connecting two macroscopic electrodes (depicted schematically
in Fig.\ 1) is an open quantum mechanical system.  The Schr\"odinger 
equation for such an open system is most naturally formulated as a scattering
problem.  The fundamental quantity
describing the properties of the system is the scattering matrix
$S(E)$ connecting incoming and outgoing asymptotic states of conduction
electrons in the electrodes.  (The degrees of freedom corresponding to the
metallic ions and core electrons will not be treated explicitly, but will be
assumed to give rise to a confinement potential for the conduction electrons,
i.e., to specify the geometry of the system.  This should be a reasonable
starting point to describe simple monovalent metals.)
The formulation of electrical transport in terms of the scattering matrix 
was developed by Landauer\cite{landauer} and B\"uttiker,\cite{condtheory}
while the formulation of the 
statistical mechanics of open quantum systems in terms of the 
scattering matrix was
first given by Dashen, Ma, and Bernstein,\cite{statmech} and was
recently revived in the context of the persistent current problem by
Akkermans {\it et al}.\cite{Iopen}  A unified treatment of the electrical
and mechanical properties of metallic nanostructures in terms of the
electronic scattering matrix was given by Stafford, Baeriswyl, 
and B\"urki.\cite{sbb}
In the remainder of this section, we recapitulate the general formalism of
Ref.\ \onlinecite{sbb}, which
will serve as the starting point for the present investigation.

The essential ingredient in the scattering matrix description of mesoscopic
systems is that electrons are injected into the system from {\em
macroscopic} reservoirs in internal thermal equilibrium; any perturbation of
the reservoirs due to the mesoscopic current flowing from one to another is
assumed to be negligible.  
The energy distribution of the electrons injected from reservoir $\alpha$
is thus given by the
Fermi distribution function $f_{\alpha}(E)=\{\exp[\beta(E-\mu_{\alpha})] 
+ 1\}^{-1}$,
where $\mu_{\alpha}$ and $\beta=1/k_B T$ are the electrochemical potential and 
inverse temperature, respectively, of reservoir $\alpha$.  

The asymptotic scattering states of conduction electrons for
the geometry depicted in Fig.\ 1 are described by a transverse
quantum number $\nu$ and a wave number $k$ which is a function of
energy $E$ and $\nu$:
\begin{equation} 
k_\nu(E)=\sqrt{\frac{ 2 m }{\hbar ^2} \left(E-\epsilon_\nu \right)}\;,
\end{equation}
where $ \epsilon_\nu$ is the energy of the transverse modes.  If the
amplitudes of incoming currents at energy $E$ are given by a vector
with components $a_\nu^+$ (currents from the left side) and $a_\nu^-$
(currents from the right side), the outgoing current amplitudes
$b_{\nu'}^+$ and $b_{\nu'}^-$ are given by
\begin{equation}
\left( 
\begin{array}{c}
b^+ \\ b^- 
\end{array} \right) =
S(E) \left( \begin{array}{c} a^+ \\ a^- \end{array}
\right).
\end{equation}
The scattering matrix for a two-terminal conductor may be decomposed into
four submatrices
\begin{equation}
S(E)=\left( \begin{array}{cc} S_{11}&S_{12}\\ S_{21}&S_{22}
\end{array} \right),
\label{s-matrix}
\end{equation}
where the submatrices $S_{ij}(E)$ contain transmission ($i \ne j$) and
reflection ($i=j$) amplitudes.   Each submatrix $S_{ij}(E)$ is a matrix
in the incoming and outgoing scattering channels $\nu$ and $\nu'$.

The electrical conductance is given in terms of the scattering matrix
by the well-known formula\cite{landauer,condtheory}
\begin{equation}
\label{Landauer_form}
G=\frac{ 2 e^2}{h} \int dE\, \frac{-\partial f(E)}{\partial E} 
{\rm{Tr}} \left\{S_{12}^\dagger(E) S_{12}(E) \right\},
\label{eq.cond}
\end{equation}
where the factor of 2 accounts for spin degeneracy.   
Eq.\ (\ref{Landauer_form})
indicates that $G$ is proportional to the sum over the transmission 
probabilities of electrons incident in a window of width $k_B T$ about 
the Fermi energy $\mu$.
Eq.\ (\ref{Landauer_form})
may be modified by electron-electron interactions at finite temperatures,
but has been shown\cite{meir_wingreen} to hold quite generally 
in the limit $T\rightarrow 0$.

The quantity needed to investigate the statistical mechanics of the 
nanowire is the electronic density of states, $D(E)$.
The density of states can be expressed in terms of the scattering matrix as
\cite{statmech,partial.dos}
\begin{equation}
\label{density_of_states}
D(E)=\frac{ 1}{2 \pi i} {\rm Tr}\left( S^\dagger(E) \frac{\partial
S(E)}{\partial E} - {\rm h.c.} \right).
\label{DOS}
\end{equation}
This expression holds for an arbitrary interacting 
gas of particles.\cite{statmech}
Given the density of states, the grand partition function may be evaluated,
and the thermal expectation values of all observables may be calculated.

The expectation value of the total electronic charge on the nanowire is
given by
\begin{equation} 
\langle Q^- \rangle = -e \int dE\, f(E) D(E).
\label{qtotal}
\end{equation}
Integrating by parts, and taking the limit $T\rightarrow 0$, one obtains the
simplified expression
\begin{equation}
\langle Q^- \rangle = -\frac{e}{\pi} \mbox{Im} \{\ln \det S(E_F)\}.
\label{qfromS}
\end{equation}
The overall phase of the scattering matrix depends on the 
exact choice of the asymptotic states (their phase). Therefore the
phase relation chosen between the amplitudes $a_{\nu}^{\pm}$ and
$b_{\nu'}^{\pm}$ is in principle free. 
Different choices correspond to the inclusion of various amounts of the 
constant asymptotic charge density in $Q^-$.  However, the total charge
$Q=\langle Q^- \rangle + Q^+$ is independent of this choice of phase, 
provided we add the appropriate quantity of the constant positive background
charge density. 

The grand canonical potential is 
the relevant thermodynamic potential to describe the mechanical properties
of the electron gas in the nanowire, and
may be written 
\begin{equation}
\label{omega_allgemeiner_Ausdruck}
\Omega =- \frac{ 1}{\beta} \int dE\, D(E) \ln(1+e^{-\beta (E-
\mu)}).
\end{equation}
It should be noted that Eq.\ (\ref{omega_allgemeiner_Ausdruck}) only holds
for noninteracting electrons, since the thermal trace is taken assuming 
each fermionic mode is independent.  A more general expression for
$\Omega$, valid within a self-consistent mean-field treatment of interactions,
will be presented elsewhere.\cite{unpub}  Here it is sufficient to note that
it will be argued below that the 
corrections to Eq.\ (\ref{omega_allgemeiner_Ausdruck}) due to charging
effects are negligible.

Eqs.\ (\ref{qtotal}) and
(\ref{omega_allgemeiner_Ausdruck}) are deceptively simple,
and it is worth emphasizing that $\mu$ is the {\em asymptotic} electrochemical
potential of electrons injected from 
the reservoirs, {\em not} a local Fermi energy, as
introduced by certain other authors.\cite{vRuitenbeek,landman2}  $D(E)$
is the global energy density of eigenstates of the scattering
problem, and contains all effects of multiple scattering, quantum
interference, 
etc.  These eigenstates are populated according to the Fermi
distributions of the reservoirs.  The occupation of a local basis of states,
which are not eigenstates of $H$, is in general quite complicated, and will be
discussed in detail in Sec.\ \ref{sec_lldos}.

In the following, it will be assumed that the volume $Ld^2$
of the nanowire (or the area $Ld$ for 2D nanowires)
remains constant under elongation, {\it i.e.,} we assume an ideal plastic
deformation.  It can be shown that relaxation of this constraint, to include
{\it e.g.}, a small elastic deformation, does not modify the mesoscopic
effects in an essential way.\cite{unpub}
The cohesive force of the nanowire is given by the derivative of the 
grand canonical potential with respect to the elongation:
\begin{equation}
F=-\frac{ \partial \Omega}{\partial L}.
\end{equation}
Differentiating the above expression for $\Omega$ and
performing a partial
integration over $E$, one obtains the general result
\begin{equation}
\label{equ_for_the_force}
F=\frac{ 1}{\pi} \int dE \, f(E) \,{\rm Im} \left\{ \frac{
\partial}{\partial L} \ln \det[S(E)] \right\}.
\label{forcefromS}
\end{equation}
We thus have relations which give us the conductance, charge, and force as a
function of the S--matrix.

\section{The S--Matrix of the WNW--geometry}
\label{sec.comp_smatr}

\begin{figure}
\epsfxsize=14cm
\epsfysize=11.9cm
\hfil
\epsffile{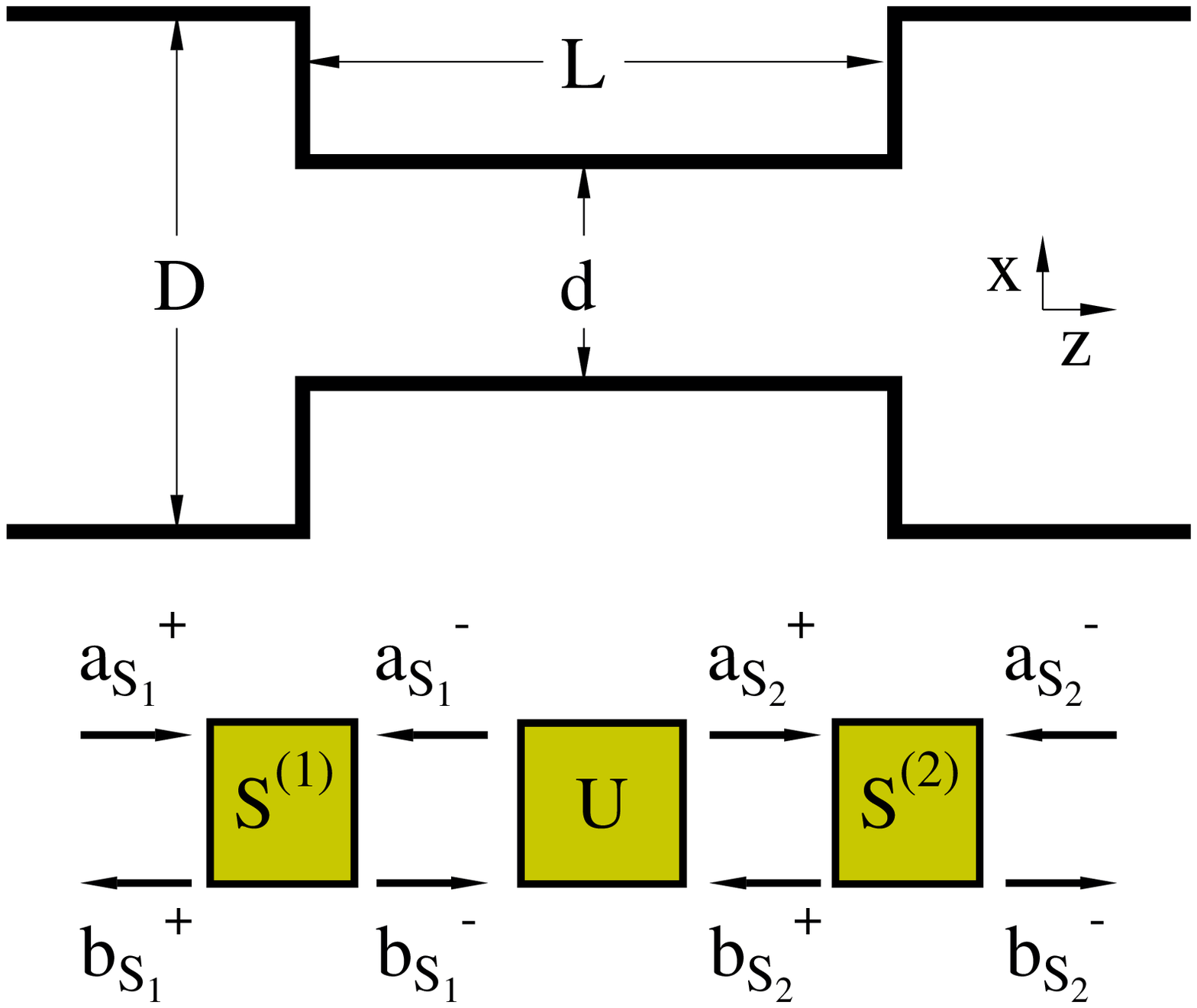}
\hfil
\caption{Schematic diagram of the WNW--Geometry. In the upper part, the
wire geometry is sketched. $D$ and $d$ are the diameters of the wide and
narrow parts of the wire, respectively.  Under elongation, it is assumed that
the area of the narrow part of the wire is conserved, $Ld=L_0 D =A$, where
$L_0$ is the initial length.
In the 3D case, the wire is assumed to have a square cross-section, and the 
volume of the narrow part is held constant during elongation, $Ld^2=L_0D^2=V$. 
The lower part of the figure shows the
scattering scheme: scattering matrices ($S^{(1)}$ and $S^{(2)}$ at
WN--transitions and $U$ for the narrow part of the wire) characterize
the transmission and reflection of current amplitudes denoted as arrows in the
diagram. The total S--matrix relates
the outgoing current amplitudes $b_{S_1}^+$ and $b_{S_2}^-$ to the
incoming current amplitudes $a_{S_1}^+$ and $a_{S_2}^-$.}
\label{fig.1}
\end{figure}

In order to obtain exact results for the tunneling and finite-size 
corrections to the conducting and thermodynamic properties of metallic
nanowires, we consider the special WNW geometry illustrated in Fig.\ 
\ref{fig.1},
for which the S--matrix can be calculated exactly.  
The system consists of a noninteracting electron gas confined by hard-wall
boundary conditions (we will return to the question of electron-electron
interactions in Sec.\ \ref{sec.screen}).
The wire cross--section in the 3D case is assumed to be
square.  A generalization to an arbitrary cross section is straightforward
if the
eigenfunctions and eigenvalues of the 2D Schr\"odinger equation are known
for that shape. The scattering problem for the 2D WNW geometry was solved by
Szafer and Stone\cite{szafer} in connection with the problem of 
conductance quantization in 2D semiconductor quantum point contacts,
and was further investigated by Weisshaar {\it et al.}\cite{weisshaar}
In addition to the transmission coefficients calculated in 
Refs.\ \onlinecite{szafer} and \onlinecite{weisshaar}, 
we need the reflection coefficients,
{\it i.e.,} the full S--matrix.
The generalization of the method of Refs.\ \onlinecite{szafer,weisshaar}
to 3D nanowires and to calculate the full S--matrix is described below.  

In order to calculate the elements of the S--Matrix, solutions of the
Schr\"odinger--equation are matched at the transitions between
wide and narrow parts of the wire.  Let us regard the transition from
wide to narrow first (cf.\ left part of Fig. \ref{fig.1}).  If the
$z$--coordinate is directed along the wire and $x$ is an abbreviation
for $x$ in the two dimensional case and for $(x_1,x_2)$ in the three
dimensional case, describing the dimension(s) perpendicular to the $z$--axis,
the wave functions are given by
\begin{eqnarray}
\Psi(z<0,x)&=&e^{i K_N z} \Phi_N(x)+ \sum_{N'} r_{N' N} e^{-i K_{N'}
z} \Phi_{N'}(x)\;,\\ \Psi(z>0,x)&=&\sum_{n} t'_{n N} e^{i k_{n} z}
\phi_{n}(x)\;,
\end{eqnarray}
where we have assumed an incoming wave from the left of unit amplitude, and
$\Phi_N$ and $\phi_n$ are transverse eigenfunctions in the wide and narrow
parts of the wire, respectively, and are given by
\begin{eqnarray}
\Phi_{N}(x)&=&\sqrt{\frac{ 2}{D}} \sin \left(\frac{N \pi}{D}(x+D/2)
\right) \;\;\;\; N=1,2,...\\ \phi_{n}(x)&=&\sqrt{\frac{ 2}{d}} \sin
\left(\frac{n \pi}{d}(x+d/2) \right) \;\;\;\; n=1,2,...
\end{eqnarray}
in the 2D case, and by an analogous expression consisting of a product
of two sine--functions for the 3D case. Here $D$ is the diameter of the 
wide part of the wire and $d$ the
diameter of the narrow part. In the 3D case, the
transverse modes have in general two indices, {\it e.g.}, $P$ and $Q$, but we
can order the states according to their energy and so characterize
them by one quantum number $N$; $P$ and $Q$ are then functions of $N$.
The wave numbers $K_N$ and $k_n$ are
\begin{eqnarray}
K_N&=&\sqrt{\frac{ 2m}{\hbar^2} E - \frac{ \pi^2 N^2}{D^2}}\;,\\
k_n&=&\sqrt{\frac{ 2m}{\hbar^2} E - \frac{ \pi^2 n^2}{d^2}}\;,
\end{eqnarray}
where we use the abbreviation (for the 3D case) $N^2=P(N)^2+Q(N)^2$.

The solution of the Schr\"{o}dinger equation must obey two
conditions at the transition point $z=0$:
\begin{enumerate}
\item
Continuity of the wave function for $x \in [-D/2,D/2]$ in 2D ($x_1,x_2 \in
[-D/2,D/2]$ in 3D):
\begin{equation}
\Phi_N(x)+\sum_{N'} r_{N'N} \Phi_{N'}(x) = \sum_n
t'_{nN}\phi_n(x) \Theta(d/2-|x|)\;,
\end{equation}
where $\Theta(x)$ is the Heavyside step function and is an
abbreviation for the product $\Theta(d/2-|x_1|) \Theta(d/2-|x_2|) $ in
the 3D case.
\item
Continuity of the first derivative of the wave function for 
$x \in [-d/2,d/2]$ in 2D ($x_1,x_2 \in [-d/2,d/2]$ in 3D):
\begin{equation}
K_N \Phi_N(x)- \sum_{N'} r_{N'N} K_{N'} \Phi_{N'}(x) = \sum_n
t'_{nN} k_n \phi_n(x) \;.
\end{equation}
\end{enumerate}
These equations can be transformed into matrix equations for $r$ and
$t'$ by multiplication with $\Phi_M(x)$ and $\phi_m(x)$,
respectively, and
integration over the appropriate $x$--range. Using the abbreviations
\begin{eqnarray}
\rho_{Nn}&=&\int_{-d/2}^{d/2} dx \, \Phi_N(x) \phi_n(x)\;,\\
K_{NN'}&=&\delta_{NN'} K_N \;,\;\; k_{nn'}\,=\,\delta_{nn'} k_n\;,
\end{eqnarray}
two equations for $r$ and $t$ are obtained:
\begin{eqnarray}
\label{matrix_equa_wn_1}
1+r&=&\rho t'\;,\\
\label{matrix_equa_wn_2}
\rho^T K - \rho^T K r &=& k t'\;.
\end{eqnarray}
Note that the $\rho$ matrix is not orthogonal.  The equations
(\ref{matrix_equa_wn_1}) and (\ref{matrix_equa_wn_2}) can be solved,
and we obtain
\begin{eqnarray}
t'&=&2(k+\rho^T K \rho ) ^{-1} \rho^T K \;,\\ r&=& \rho t'
- 1\;.
\end{eqnarray}
An exactly analogous calculation for an incoming wave from the right
side gives
\begin{eqnarray}
r'&=& (k+\rho^T K \rho)^{-1} (k-\rho^T K \rho)\;,\\ t&=&
\rho+ \rho r'\;.
\end{eqnarray}
The scattering matrix is obtained by normalizing the wave
amplitudes with respect to current (the unitarity of $S$ reflects current
conservation). With
\begin{equation} 
\bar{t}_{nN}=(k_n/K_N)^{1/2}
t_{nN}\,,\;\;\bar{r}_{NN'}=(K_N/K_{N'})^{1/2} r_{NN'}\,,\;\;\mbox{etc.,}
\end{equation}
the scattering matrix is given by
\begin{equation} 
S^{(1)}=\left(\begin{array}{cc} \bar{r} & \bar{t} \\ \bar{t}' &
\bar{r}' \end{array} \right)\;.
\label{smatrix.1}
\end{equation}

The S--matrix for the combined WNW--geometry may be constructed from 
three scattering matrices $S^{(1)}$, $U$, and $S^{(2)}$, describing the 
scattering at the WN boundary, the free propagation within the narrow 
section, and the scattering at the NW boundary, respectively 
(see Fig.\ \ref{fig.1}). 
The free propagation in the narrow section is described by the matrix
\begin{equation}
U=\left( \begin{array}{cc} 0 & X \\ X & 0 \end{array} \right)\;;\;
X_{n n'}=\delta_{n n'} \exp(i k_n L)\;.
\label{smatrix.free}
\end{equation}
The NW transition is associated with a matrix
\begin{equation}
S^{(2)}=\left(\begin{array}{cc} \bar{r}' & \bar{t}' \\ \bar{t} &
\bar{r} \end{array} \right)\;,
\label{smatrix.2}
\end{equation}
which can be calculated like $S^{(1)}$, or can be seen by symmetry
considerations.  To compute the full S--matrix, we use the linear equations
connecting the current amplitudes 
propagating between the individual scattering matrices (see
Fig.\ \ref{fig.1} for an explanation of the notation):
\begin{eqnarray}
\left(\begin{array}{c}b_{S_1}^+ \\ b_{S_1}^- \end{array} \right) &=&
S^{(1)} \left(\begin{array}{c}a_{S_1}^+ \\ a_{S_1}^- \end{array}
\right)\;,\nonumber\\ \left(\begin{array}{c}a_{S_1}^- \\ a_{S_2}^+ \end{array}
\right) &=& U \left(\begin{array}{c}b_{S_1}^- \\ b_{S_2}^+ \end{array}
\right)\;, \label{submatrix}\\ 
\left(\begin{array}{c}b_{S_2}^+ \\ b_{S_2}^- \end{array}
\right) &=& S^{(2)} \left(\begin{array}{c}a_{S_2}^+ \\ a_{S_2}^-
\end{array} \right)\;.\nonumber
\end{eqnarray}
Eliminating the unwanted variables in this set of linear equations, and
rewriting the equations in the form
\begin{equation}
\left(\begin{array}{c}b_{S_1}^+ \\ b_{S_2}^- \end{array} \right) = S
\left(\begin{array}{c}a_{S_1}^+ \\ a_{S_2}^- \end{array} \right),
\end{equation}
relating ingoing and outgoing currents, the full S--matrix is found to be
\begin{equation}
\label{WNW_S_Matrix}
S= P \left(\begin{array}{cc} S_{11}^{(1)}+S_{12}^{(1)}
(1-U_{12}S_{11}^{(2)} U_{21} S_{22}^{(1)} )^{-1} U_{12} S^{(2)}_{11}
U_{21} S^{(1)}_{21} & S_{12}^{(1)} (1- U_{12}S_{11}^{(2)} U_{21}
S_{22}^{(1)} )^{-1} U_{12} S_{12}^{(2)} \\ \\ S_{21}^{(2)} (1-
U_{21}S_{22}^{(1)} U_{12} S_{11}^{(2)} )^{-1} U_{21} S_{21}^{(1)} &
S_{22}^{(2)}+S_{21}^{(2)} (1-U_{21}S_{22}^{(1)} U_{12} S_{11}^{(2)}
)^{-1} U_{21} S^{(1)}_{22} U_{12} S^{(2)}_{12}
\end{array} \right)P.
\label{Stotal}
\end{equation}
The operator $P$ is a projection operator onto the 
undamped modes in the wide part of the wire.   Note that the individual
matrices $S^{(1)}$, $S^{(2)}$, 
and $U$ are infinite-dimensional matrices describing
scattering and propagation in all available modes, including the evanescent
modes.  The full S--matrix, on the other hand, connects the incoming and
outgoing asymptotic states, and thus has a finite dimension for a given
energy $E$, determined by the total number of transverse modes $\nu$ with 
$\epsilon_{\nu} \leq E$.  The inclusion of the virtual intermediate 
states, which describe tunneling processes,
is crucial to solve the Schr\"odinger equation accurately. 
However, the contribution of the evanescent modes decreases exponentially 
with increasing energy.
In practice, we found numerical convergence of the S--matrix 
if roughly 20 times more modes than the undamped modes in the
W--part were retained.

We remark that
the elements of the S--matrix can also be found by considering
{\it e.g.}, the transmission matrix $S_{1 2}$ 
as the sum of the directly transmitted current amplitudes
and the multiply backscattered current amplitudes.  This results in
a geometric series, which can be summed to obtain the result
(\ref{Stotal}).

\section{2D Nanowire}
\label{sec.2Dnanowire}

In this section, we investigate the properties of 2D nanowires.  
There are several motivations to study 2D systems:  First, a quasi-2D 
nanowire could be experimentally realized in a thin metallic film.  Secondly,
the characteristic electrical and mechanical properties of a nanowire, namely,
conductance quantization and force oscillations, are already present in 
2D systems, and 
it is worthwhile to investigate to what extent the universality of the 
mesoscopic force oscillations predicted in Ref.\ \onlinecite{sbb} depends
on dimensionality.  Further, 2D electronic structure is 
easier to visualize, making it simple to study the correlations between the 
measured quantities and the local electronic structure.  
Finally, and perhaps most importantly, certain of the 
phenomena studied here are directly applicable to 2D quantum point contacts
formed in semiconductor heterostructures.\cite{vanwees,wharam,beenakker}
While the predicted mesoscopic force oscillations of order $E_F/\lambda_F$
would be many orders of magnitude smaller in doped semiconductors due to the 
smaller Fermi energy and correspondingly longer Fermi wavelength, and would
likely be hidden by the much larger cohesive forces associated with the 
covalently bonded electrons of the valence band, the charge
oscillations predicted to accompany the quantized steps in the conductance
should have a comparable size in both metallic and semiconductor quantum
point contacts, namely, of order the fundamental charge quantum $e$.

\subsection{Force and conductance}
\label{force_and_cond}

\begin{figure}
\epsfxsize=14cm
\epsfysize=10.5cm
\hfil
\epsffile{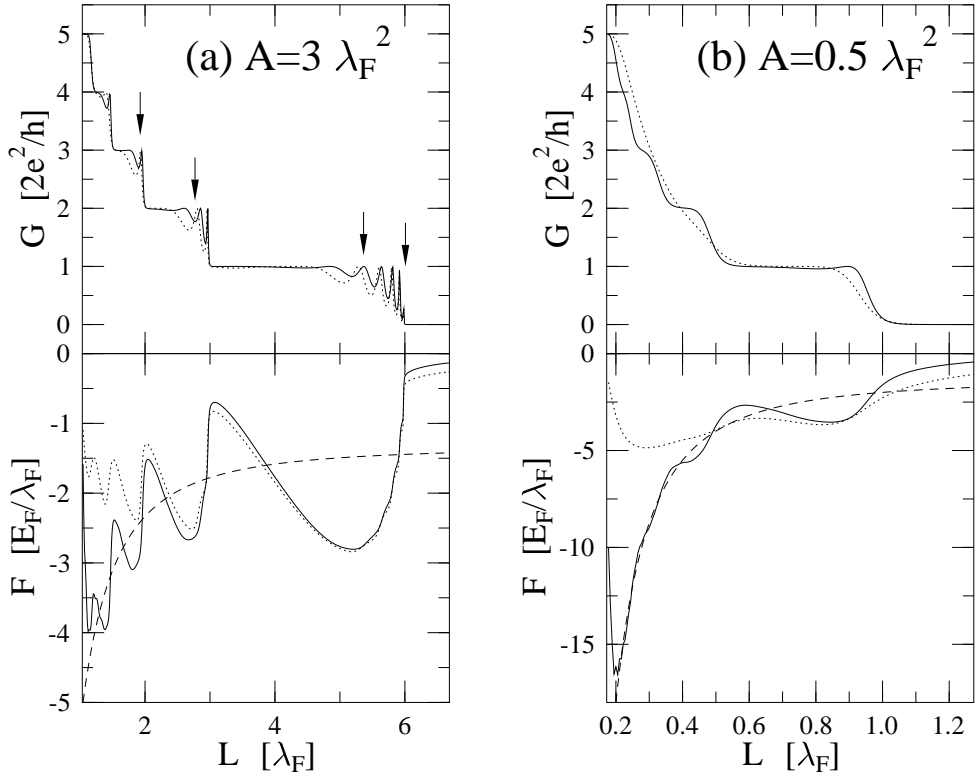}
\hfil
\caption{Electrical conductance and tensile force for two different 2D
wires with WNW--geometry as a function of the length $L$ of the narrow
part. Assuming area conservation, $L$ is varied from a perfect wire
(where narrow and wide parts have the same diameter) until the last
conductance channel breaks. (a) A wire with a relatively large area of
the narrow part ($A=3.0 \lambda_F^2$) and (b) a wire with a smaller area
($A=0.5 \lambda_F^2$) are shown. The diameter of the wide part is $2.9
\lambda_F$, fixing the total number of asymptotically propagating modes to be 
5. The dotted curves show the adiabatic approximation
to conductance and force, the dashed curves give the surface tension,
the leading order contribution in a semiclassical expansion of the force.
The arrows indicate the geometries used in
Fig.\ \ref{fig.4}.}
\label{fig.2}
\end{figure}

Once the S--matrix (\ref{WNW_S_Matrix}) is known, the conductance and
cohesive force can be calculated from Eqs.\ (\ref{Landauer_form})
and (\ref{equ_for_the_force}). 
Fig.\ \ref{fig.2} shows the behavior of the conductance and cohesive force
as a function of the elongation of the wire. An ideal plastic deformation of
the narrow part is assumed, which means that its area $A=Ld=L_0 D$ is held
constant, $L_0$ being the initial length of the narrow section.  
The wide sections of the wire support 5 propagating modes at the
Fermi energy; this fixes the conductance of the wire before deformation to
be $G=5G_0$.  In
Fig.\ \ref{fig.2}(a), the area of the N--part is comparatively large ($3
\lambda_F^2$), while it is small in Fig.\  \ref{fig.2}(b) ($0.5
\lambda_F^2$).  In both cases, conductance quantization can be
observed; whenever a channel in the N--part closes (as the diameter is
decreased), the conductance decreases and reaches a plateau at an
integer multiple of $2 e^2/h$.  The conductance does not show a
perfect step--like structure, though:  When the narrow part is short,
there will be tunneling through the constriction before a channel
opens, and reflection above the threshold.  This will smear out
the steps, leading to a rather smooth transition between the plateaus
[see Fig.\ \ref{fig.2}(b)]. When the narrow part is quite long, on the
other hand, there is almost no tunneling, but a resonant structure
near the transition points occurs [Fig.\ \ref{fig.2}(a)]. This is due to the 
alternating constructive and destructive internal reflection within the
constriction.\cite{szafer}

The cohesive force is strongly correlated with the conductance. 
In Fig.\ \ref{fig.2}(a), 
the modulus of the force increases along the conductance plateaus, while it
decreases sharply at the conductance steps.  The behavior is qualitatively 
similar to the result for smooth 3D 
geometries presented in Ref.\ \onlinecite{sbb}, and to the experimental results
for 3D Au nanowires.\cite{nanoforce,nanoforce2}  Thus we see that the 
essential
correlations of the electrical and mechanical properties of nanowires 
are present even in 2D systems, and even for abrupt geometries.
For the extremely short nanowire considered in Fig.\ \ref{fig.2}(b), similar
force oscillations correlated with the conductance steps are present
(see Fig.\ \ref{figq1}), but they
are superimposed on a much larger background force.
The pronounced difference in the force in Figs.\ \ref{fig.2}(a) and 
\ref{fig.2}(b) indicates a breakdown of the invariance of $F$ 
under a stretching of the geometry $d(z)\rightarrow d(\lambda z)$, which
was found within the WKB approximation,\cite{sbb} due to strong tunneling 
effects in very short wires. 

The cohesive force decays to zero as the wire is elongated past the point where
the last conductance channel is cut off, though rather more slowly than the 
conductance itself.  $F$ remains noticably finite even for the largest 
elongations shown in Figs.\ \ref{fig.2}(a) and \ref{fig.2}(b), although 
$G$ is exponentially small.\cite{remark.jerome} The force in this regime
arises from the variation of the free energy due to a deformation of the 
geometry in the region classically forbidden to electrons, and is clearly larger
in the shorter wire [Fig.\ \ref{fig.2}(b)], 
where tunneling effects are more important.  This effect is simple to 
understand:  even when the probability to tunnel all the way through the 
narrow section is exponentially small, the probability to {\em enter} the
narrow section need not be small [see Fig.\ \ref{fig.4}(d)], so the electron
gas is still sensitive to its shape.

It is clear from Eq.\ (\ref{forcefromS}) that all states with
energy smaller than the Fermi energy contribute to the total
force. On the other hand, the
graphs show that the force oscillations are correlated to the behavior of
the conductance, and thus must be due to states near the
Fermi energy.  In Fig.\ \ref{fig.2}(a), one can see that even the
resonant structure in the conductance is reflected in the force,
leading to sudden changes in its derivative.

\begin{figure}
\epsfxsize=7cm
\epsfysize=10.5cm
\hfil
\epsffile{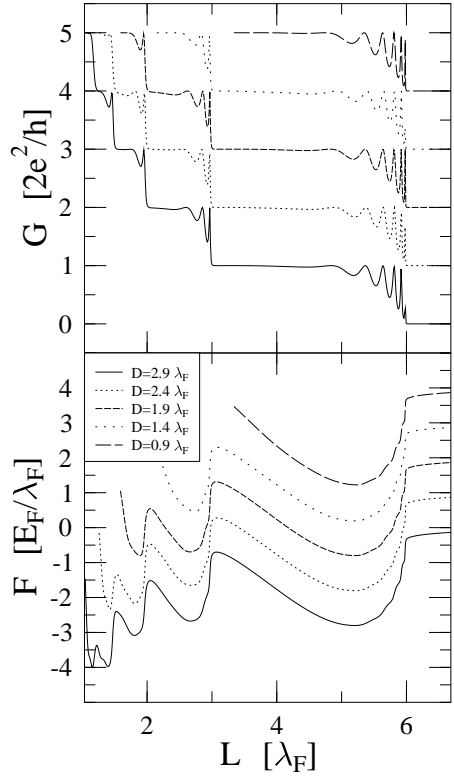}
\hfil
\caption{Electrical conductance and cohesive force for 2D wires with
various outer diameters $D$. The area of the narrow part was
held constant, so the initial length (when $D=d$) of the wires is
different. The area of the narrow part is $3.0 \lambda_F^2$ as in
Fig.\ \ref{fig.2}(a). The curves are vertically offset.}
\label{fig.3}
\end{figure}

Fig. \ref{fig.3} shows the conductance and cohesive force as a function of
elongation for wires with different outer diameters.
While the curves are distinct at the beginning of the
elongation (where inner and outer parts have comparable diameters),
there is almost no difference at higher elongations, and this is not
only valid for the conductance \cite{weisshaar} but also for the
force. This shows that even a narrow constriction coupled to contacts
with infinite width can be accurately modeled by wires of the type we
regard here.

\subsection{Mesoscopic force and charge fluctuations}
\label{sec.osc}

In order to understand the overall behavior of the cohesive force, it is 
useful to consider a semiclassical expansion\cite{sbb,commentzwerger} of Eq.\ 
(\ref{omega_allgemeiner_Ausdruck}).  Formally, the density of states 
may be written $D(E) = dN(E)/dE$,
where $N(E)$ is the total number of states with energy less
than $E$ in the system.  
The behavior of $N(E)$ for 2D domains with hard-wall boundary
conditions was first investigated by Weyl,\cite{weyl} and was further
developed by Kac and others.\cite{kac,corners}
The Weyl expansion of $N(E)$ for a 2D domain with a polygonal boundary with $n$
corners is\cite{corners}
\begin{equation}
N(E)= \frac{A}{2\pi}k_E^2 - \frac{\partial A}{2\pi} k_E + 2C + \delta N(E),
\label{eq.weyl}
\end{equation}
where $k_E = \sqrt{2mE/\hbar^2}$ is the wavevector associated with energy
$E$, $A$ is the area of the domain, $\partial A$
the circumference of the domain, and
C is a constant depending on the topology, in this case
\begin{equation}
C=\sum_{i=1}^n \frac{\pi^2-\gamma_i^2}{24\pi\gamma_i},
\label{top.term}
\end{equation}
where $\gamma_i$ is the interior angle of corner $i$, and $\delta N(E)$
is a fluctuating term associated with the discreteness of the level
spectrum, whose energy average is zero.\cite{commentzwerger}
Note that Eq.\ (\ref{eq.weyl}) includes a factor of 2 for spin.
The shift in the total number of modes due to the presence of the sharp 
corners is given by $C=1/9$ for the WNW geometry of Fig.\ \ref{fig.1}.

Integrating Eq.\ (\ref{omega_allgemeiner_Ausdruck}) by parts and taking 
the limit $T\rightarrow 0$, one finds
\begin{equation}
\Omega= -\int_0^{E_F} N(E)\, dE = -\frac{\pi E_F}{\lambda_F^2} A
+ \frac{2E_F}{3\lambda_F} \partial A - 2 E_F C + \delta \Omega,
\label{omega.weyl}
\end{equation}
where $\delta \Omega$ is a fluctuating mesoscopic correction.
Differentiating $\Omega$ with respect to $L$, subject to the constraint
$A=\mbox{const.}$, yields a semiclassical expansion for the force
\begin{equation}
F= -\frac{2E_F}{3\lambda_F} \frac{d}{dL}\partial A + \delta F,
\label{f.weyl}
\end{equation}
where $\delta F=-\partial (\delta \Omega)/\partial L$.
The leading order term in the semiclassical
expansion of the force is the surface tension, $F_{\rm surf}$. 
For the WNW geometry, the circumference of the nanowire is
$\partial A = 2 L + 2(D-d) + \mbox{const.}$, and one obtains
\begin{equation}
F_{\rm surf} = - \frac{4E_F}{3\lambda_F} \left( 1 +
\frac{d^2}{A}\right).
\label{F.surf.WNW}
\end{equation}
This indicates that the surface tension increases with increasing conductance
(the Sharvin formula reads $G /G_0 \sim 2d/\lambda_F$ in 2D) 
and with {\it decreasing area} of the wire.
The surface tension is plotted in Fig.\ \ref{fig.2} as a dashed
curve. The exact force oscillates around it.  

\begin{figure}
\epsfxsize=14cm
\epsfysize=10.5cm
\hfil
\epsffile{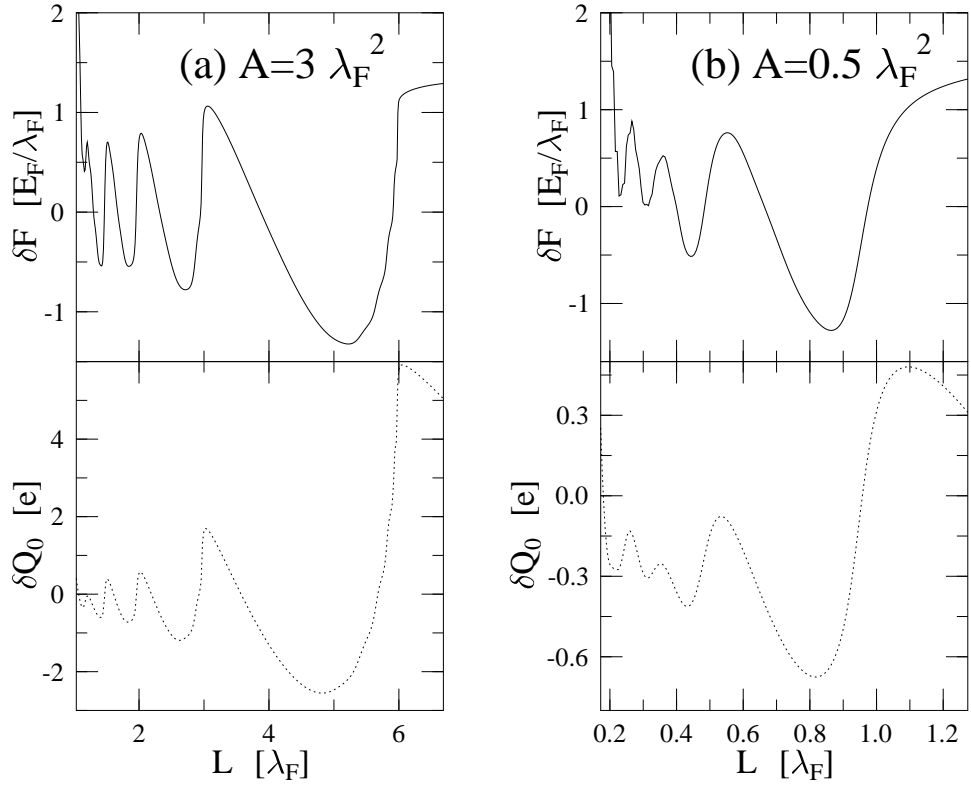}
\hfil
\caption{Force and charge oscillations for two different 2D WNW
wires as a function of elongation. Force oscillations are shown in the
upper and charge oscillations in the lower half. The wire parameters are the same as in the preceding figures, 
(a) shows a wire with a larger and (b) a wire with a smaller area of
the narrow part.}
\label{figq1}
\end{figure}

The Weyl expansion for the electronic charge of the nanowire for $T
\rightarrow 0$ is
\begin{equation}
\langle Q^-\rangle = -e N(E_F) = -e \left( \frac{2\pi A}{\lambda_F^2}
-\frac{\partial A}{\lambda_F} + 2C\right) + \delta Q_0.
\label{weyl.charge}
\end{equation}
In Fig.\ \ref{figq1}, the force fluctuations $\delta F$, calculated by 
subtracting the surface tension from the exact force, and the charge
fluctuations $\delta Q_0$, calculated by subtracting the term in parentheses
in Eq.\ (\ref{weyl.charge})
from the exact charge, determined from Eq.\ (\ref{qfromS}), are shown for the
nanowires considered in Fig.\ \ref{fig.2}.
The scale of the force
oscillations is $E_F/\lambda_F$ (see Fig.\ \ref{figq1}), 
similar to the result for smooth 3D 
geometries presented in Ref.\ \onlinecite{sbb}. 
Strongly correlated with the force oscillations are charge oscillations
of order the fundamental charge quantum $e$.
The charge fluctuations $\delta Q_0$ are calculated in the absence of screening.
Screening will be considered in Sec.\ \ref{sec.screen};  
here it suffices to note 
that the screening properties of the 2D electron gas in {\it e.g.}
GaAs are quite poor, so that the charge fluctuations in small-conductance
2D quantum point contacts should be essentially unscreened.  The predicted 
charge oscillations should be experimentally observable with a local probe,
such as a single-electron transistor.

\subsection{Topological force}
\label{sec.top}

A close examination of Fig.\ \ref{fig.2} indicates that the exact 
force deviates significantly from the surface tension for very small
elongations:
The discrepancy is roughly $3 E_F/\lambda_F$
in Fig.\ \ref{fig.2}(a) and $14 E_F/\lambda_F$
in Fig.\ \ref{fig.2}(b) for $L\rightarrow L_0$.  
This behavior is to be contrasted with the 
results for smooth geometries, presented in Ref.\ \onlinecite{sbb},
in which the mesoscopic deviations from the semiclassical result were always
found to be $\lesssim E_F/\lambda_F$ (hence the term {\it universal}).  
The nonuniversal corrections to the force in the WNW geometry at small
deformations have a topological origin:  Before deformation, the perfect 
wire has a smooth boundary, but as the wire is stretched, eight sharp
corners develop.  The sharp corners lead to a shift in the grand canonical
potential 
\begin{equation}
\Delta \Omega_{\rm top} = - 2 E_F C = -\frac{2}{9}E_F.
\label{omega.top}
\end{equation}
However, the 
electrons incident from the reservoirs can only resolve the individual
corners when they are separated by a distance greater than or
of order $\lambda_F$.  Thus one may expect the topological correction to 
$\Omega$ due to the sharp corners to evolve smoothly with the initial
deformation, and to saturate when
the corners become separated by an amount of order $\lambda_F$.
In order to see this explicitly, let us consider $\Delta \Omega_{\rm top}$
as the work done by a topological force $F_{\rm top}=F-F_{\rm surf}$ which
is present for small deformations $L\approx L_0$.   For infinitessimal
deformations, the adiabatic approximation becomes exact, and gives
$F(L_0)=-4E_F/3\lambda_F$.  Thus $F_{\rm top}(L_0)=4E_F D/3\lambda_F L_0$.
Estimating the work done by $F_{\rm top}$ to be -$F_{\rm top}(L_0)
\Delta L/2$, and equating this to the change in the grand canonical potential
$\Delta \Omega_{\rm top}$, gives $\Delta L = \lambda_F L_0/3D$ as the
deformation regime where the topological force is important.  The
corresponding change in diameter is 
\begin{equation}
\Delta d = -D\Delta L/L_0 = - \lambda_F/3,
\label{diam.top}
\end{equation}
indicating that the shift in the free energy of the system due to the 
introduction of sharp corners indeed saturates when the separation between
the corners becomes comparable to the Fermi wavelength.
That the force associated with this change of topology can be large compared
to $E_F/\lambda_F$ is a remarkable result.

\subsection{Comparison with the adiabatic approximation}
\label{sec.adiabat}

In previous theoretical investigations of the conductance and cohesive force
in metallic nanowires, an adiabatic approximation\cite{sbb} was employed.
The WNW geometry clearly violates the conditions of validity of the adiabatic
approximation;  it is nonetheless instructive to compare our exact results
to those obtained within an adiabatic approximation for this abrupt geomety,
in order to evaluate the importance of interchannel scattering.

The solution of the 2D Schr\"odinger equation can be written formally as
$\Psi(x,z)=\phi_z(x) \psi(z)$.
The adiabatic approximation consists of neglecting 
the derivatives $\partial \phi_z(x)/\partial z$ and
$\partial^2 \phi_z(x)/\partial z^2$ 
(which would be justified if $|\partial D(z)/\partial z| \ll 1$), 
so that the Schr\"odinger equation decouples into separate transverse 
and longitudinale wave equations,
\begin{equation}
-\frac{\hbar^2}{2m} \frac{\partial^2}{\partial x^2} \phi_z^{\nu}(x)
=E_{\nu}(z) \phi_z^{\nu}(x),
\label{eq.trans}
\end{equation}
\begin{equation} 
-\frac{\hbar^2}{2 m} \frac{ \partial^2}{\partial z^2}
\psi(z)=\left(E-E_\nu(z)\right) \psi(z).
\end{equation}
The 2D scattering problem then decouples into 
independent one-dimensional scattering problems for each channel,
and the scattering matrix reduces to a $2 \times 2$ matrix for
each channel, which can be computed {\it e.g.}, via the WKB 
approximation.\cite{sbb} 
It is then straightforward to calculate conductance and
cohesive force using the formalism described in Sec.\ \ref{sec.smatrix}.

While the adiabatic approximation should be a good approximation for
boundaries with smoothly varying diameter, this condition 
is certainly not fulfilled in the WNW--geometry.  Employing it
nonetheless, the equation of motion for the longitudinal coordinate
becomes just the 1D Schr\"odinger equation for a square barrier:
\begin{equation} 
\left(-\frac{\hbar^2}{2 m} \frac{ \partial^2}{\partial z^2}+\theta(z)
\theta(L-z) V_\nu \right) \phi(z) = \tilde{E}_\nu \phi(z)
\end{equation}
with $V_\nu=E_F \frac{\pi^2 \nu^2}{k_F^2} \left(\frac{1}{d^2}-\frac{
1}{D^2} \right)$ and $\tilde{E}_\nu=E-E_F \frac{ \pi^2 \nu^2}{k_F^2
D^2}$.  The transmission and reflection coefficients are simply calculated
using the continuity of the wave function and its derivative at the
potential steps.  We find
\begin{eqnarray} 
r_\nu&=&\frac{\left( 1 - {e^{2 i k_\nu L}} \right) \left( {k_\nu^2} -
           {K_\nu^2} \right) }{-{k_\nu^2} + {e^{2 i k_\nu L}}
           {k_\nu^2} - 2 k_\nu K_\nu - 2 {e^{2 i k_\nu L}} k_\nu K_\nu
           - {K_\nu^2} + {e^{2 i k_\nu L}} {K_\nu^2}}\;,
\nonumber\\
           t_\nu&=&{{\frac{-4 {e^{i k_\nu L}} k_\nu K_\nu} {-{k_\nu^2}
           + {e^{2 i k_\nu L}} {k_\nu^2} - 2 k_\nu K_\nu - 2 {e^{2 i
           k_\nu L}} k_\nu K_\nu - {K_\nu^2} + {e^{2 i k_\nu L}}
           {K_\nu^2}}}}\;,
\label{tandr.ad}
\end{eqnarray}
where $K_\nu = k_F \sqrt{\tilde{E}_\nu/E_F}$ and $k_\nu = k_F
\sqrt{(\tilde{E}_\nu-V_\nu)/E_F}$.

Inserting the S--matrix elements (\ref{tandr.ad}) into Eqs.\ (\ref{eq.cond})
and (\ref{forcefromS}), one obtains the adiabatic approximation for the 
conductance and force.  These are shown as dotted curves in Fig.\ \ref{fig.2}.
The adiabatic approximation captures some of the qualitative features of the 
exact solution (solid curves), but is not 
quantitatively correct.  Since the conductance is more or less quantized, the
discrepancy with respect to the exact solution has to be small.  
Not so for the force; the adiabatic approximation clearly fails to describe
correctly even the leading order contribution to the force, the surface
tension, when several channels are transmitted, especially for very short
nanowires.  Interestingly, the adiabatic approximation {\it overestimates}
the effects of tunneling: both the conductance plateaus and the force 
oscillations are better defined in the exact calculation than in the 
adiabatic approximation.  Perhaps the most striking conclusion that one 
should draw from Fig.\ \ref{fig.3} is that even for the worst-case scenario 
of an abrupt geometry, the adiabatic approximation works remarkably well
for nanowires of length $L \gg \lambda_F$.

\subsection{Local density of states}
\label{sec_lldos}

The correlations between the quantized steps in the conductance and the
oscillations of order $E_F/\lambda_F$ in the force were interpreted in 
Ref.\ \onlinecite{sbb} in terms of a simple physical picture, which was 
essentially the converse of the conventional interpretation.
The conventional interpretation\cite{landman,todorov,mol.dyn} 
of the experiments of Refs.\ \onlinecite{nanoforce,nanoforce2}
is that the jumps in the conductance are due to abrupt changes of the 
structure of the nanowire at the atomic level, {\it e.g.}, through  the
breaking of bonds, and that these structural rearrangements manifested
themselves as abrupt changes in the cohesive force.  While certainly a 
plausible viewpoint, the strong statistical evidence
\cite{olesen,krans,histogram} for {\it conductance quantization} has
no natural explanation within this framework.  In order to substantiate
the converse point of view,
that the conductance channels themselves can be interpreted as mesoscopic bonds 
providing the cohesion, it is worthwhile investigating the local electronic
structure of a nanowire within the jellium model.

The electrical conductance 
is determined by the electronic structure of the nanowire in the vicinity
of the Fermi energy.  While the total force and charge clearly depend on
all the states with energy below $E_F$ [{\it c.f.}, Eqs.\ (\ref{qtotal})
and (\ref{forcefromS})], the mesoscopic force and charge oscillations, 
because they are correlated with the conductance steps, must also be 
essentially determined
by the electronic states at the Fermi energy.  The spatial character of
the electronic wavefunctions associated with scattering states of a
given energy $E$ is naturally represented through the local density of
states (LDOS), $D(E,x,z)$.  The LDOS of 2D quantum point contacts with 
smooth boundaries has recently been investigated by Ulreich and Zwerger.
\cite{LDOS}  

While Refs.\ \onlinecite{vRuitenbeek,landman2,blom} regarded the density
of states to be independent of the $z$--coordinate and to depend only
on the $x$--coordinate in the narrow part of the wire, this is certainly
not the case for the full solution of the problem.  The $z$--dependence
of the LDOS will be especially important when not
only the width but also the length of the nanowire are on the nanometer
scale, as in the experimentally relevant geometry. 
Since we know the
exact S--matrix and the eigenfunctions for the WNW geometry,
we can calculate the LDOS. 
The LDOS is obtained as a sum of the densities created by the
different incoming channels. Only reflected or transmitted waves from
the same channel will superpose coherently. In the wide section on the
left side of the constriction, the LDOS at the Fermi energy is given by
\begin{eqnarray}
D(E,x,z<0)= \frac{2}{h} \sum_{N=1}^{N_{\rm max}} & &
\left\{
\left|\frac{ e^{iK_Nz}}{\sqrt{\hbar K_N/m}} \Phi_N(x) + 
\sum_{N'=1}^{\infty}\left(S_{11}\right)_{N'N}
\frac{ e^{-iK_{N'}z}}{\sqrt{\hbar K_{N'}/m}} \Phi_{N'}(x) \right|^2\right.
\nonumber\\ 
&&\;\;\;+\left.\left|\sum_{N'=1}^{\infty} \left(S_{12}\right)_{N'N}
\frac{ e^{-iK_{N'}z}}{\sqrt{\hbar K_{N'}/m}} \Phi_{N'}(x) \right|^2\right\},
\end{eqnarray}
where $S_{ij}$ are the submatrices of the S--matrix (\ref{s-matrix}),
and we have chosen $z=0$ at the boundary between wide and narrow parts.
For the calculation of the LDOS in the narrow part of the wire, we
need the current amplitudes $b^-_{S_{1}}$ and $b^+_{S_{ 2 }}$ in the narrow
section as a function of
the incoming current amplitudes $a^+_{S_{ 1 }}$ and $a^-_{S_{ 2 }}$ (see
Fig.\ \ref{fig.1}). In the system of linear equations
(\ref{submatrix}),
the unwanted variables have to be eliminated, and the remaining 
equations may be rewritten as
\begin{equation}
\left(\begin{array}{c} b^-_{S_{ 1 }}\\ b^+_{S_{ 2 }} \end{array}
\right) = 
t \left(\begin{array}{c} a^+_{S_{ 1 }}\\a^-_{S_{ 2 }} \end{array}
\right) \;.
\end{equation}
We find
\begin{equation}
t=\left( \begin{array}{cc} t_{11} & t_{12} \\ t_{21} & t_{22}
	 \end{array} \right)
\end{equation}
with the components of $t$ given by the known matrices $S^{(1)}$
(\ref{smatrix.1}), $S^{(2)}$ (\ref{smatrix.2}) and $U$ (\ref{smatrix.free}) as 
\begin{eqnarray}
 t_{11} &=& \left(1-S_{22}^{(1)} U_{12} S_{11}^{(2)}U_{21}
\right)^{-1} S_{21}^{(1)}\;,
\\
 t_{12} &=&S_{22}^{(1)}U_{12} \left(1-S_{11}^{(2)} U_{21}
S_{22}^{(1)}U_{12} \right)^{-1} S_{12}^{(2)}\;,
\\
 t_{21} &=&  \left(1-S_{11}^{(2)} U_{21} S_{22}^{(1)}U_{12}
\right)^{-1} S_{11}^{(2)} U_{21} S_{21}^{(1)}\;,
\\
 t_{22} &=&  \left(1-S_{11}^{(2)} U_{21}
S_{22}^{(1)}U_{12} \right)^{-1} S_{12}^{(2)}\;.
\end{eqnarray}
The LDOS in the narrow part of the wire is obtained as 
\begin{eqnarray}
D(E,x,0<z<L) = \frac{2}{h}
\sum_{N=1}^{N_{\rm max}} & & \left\{
\left|\sum_{n'=1}^{\infty}\left[
\left(t_{11}\right)_{n'N} 
\frac{ e^{i k_{n'} z}}{\sqrt{\hbar k_{n'}/m}} \phi_{n'}(x) 
+ \left(t_{21}\right)_{n'N}
\frac{e^{-ik_{n'}(z-L)}}{\sqrt{\hbar k_{n'}/m}} \phi_{n'}(x)\right] 
\right|^2 \right.
\nonumber\\
&&\;\;\;+\left.\left|\sum_{n'=1}^{\infty} \left[
\left(t_{12}\right)_{n'N} 
\frac{ e^{i k_{n'} z}}{\sqrt{\hbar k_{n'}/m}} \phi_{n'}(x) 
+ \left(t_{22}\right)_{n'N}
\frac{ e^{-ik_{n'}(z-L)}}{\sqrt{\hbar k_{n'}/m}}\phi_{n'}(x)\right]
\right|^2\right\}.
\end{eqnarray}
The LDOS is of course symmetric about the axis $z=L/2$.

\begin{figure}
\epsfxsize=14cm
\epsfysize=16.33cm
\hfil
\epsffile{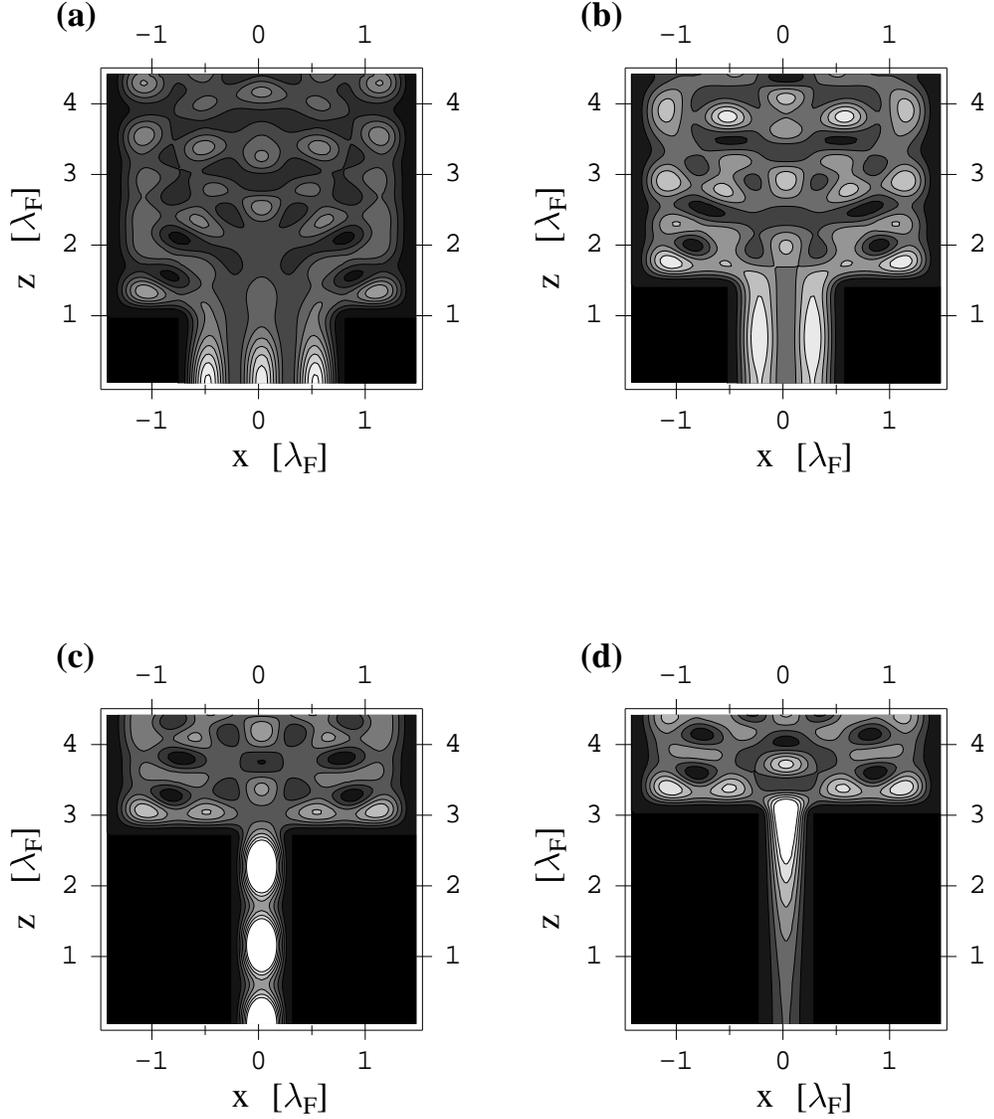}
\hfil
\caption{Local density of states $D(E_F,x,z)$ at the Fermi energy in a 2D
wire. The area of the N--part and the diameter of the W--part have
been chosen as in Fig.\ \ref{fig.2}(a) ($A=3.0 \lambda_F^2$ and $D=2.9
\lambda_F$) to make comparison with Fig.\ \ref{fig.2}(a) possible. The
diameter of the N--part is (a) $2.4 \lambda_F$, (b) $1.9 \lambda_F$,
(c) $1.4 \lambda_F$, (d) $0.9 \lambda_F$. The corresponding elongations
are marked
as arrows in the upper part of Fig.\ \ref{fig.2}(a). Black areas
correspond to $D=0.0 /(E_F \lambda_F^2)$, white areas to
$D>1.0/(E_F \lambda_F^2)$, contours are drawn at equally spaced
values of $D$ between these two limits.}
\label{fig.4}
\end{figure}

Fig. \ref{fig.4} shows the LDOS at the 
Fermi energy for 4 different elongations of a wire with a relatively
long N-part (it has the same parameters as the wire in
Fig.\ \ref{fig.2}(a), so one can compare the conductance and force at these
elongations with the LDOS). It can be seen
immediately that the LDOS exhibits a highly nontrivial structure.
The number of maxima in $x$--direction in the N--part of the wire 
reflects the number of open channels transmitted through the
constriction; in Fig.\ \ref{fig.4}(a) there are three open channels (and
thus $G \approx 3 G_0 $ see fig. \ref{fig.2}), in (b) there
are two open channels and in (c) only one channel is
left. Fig.\ \ref{fig.4}(d) shows the exponential damping of the wave
function just after the last channel has closed. So the highest open
mode dominates the transverse structure in the LDOS 
in the narrow part of the wire;
this can be understood considering that normalization of the
wavefunctions to unit current implies that the wavefunctions are
proportional to $1/\sqrt{d-d_n}$, where $d_n$ is the wire diameter at
which the $n$th channel opens.   
Note that even in the tunneling regime [Fig.\ \ref{fig.4}(d)], the probability
for an electron to enter the classically forbidden region can be non-negligible.

As the conductance is a property of states at the Fermi energy, we
should expect not only the number of transmitted channels to be
reflected in the LDOS, but also the resonant structure exhibited by
the conductance in Fig.\ \ref{fig.2}(a). This is indeed the case; when
we are at conductance maxima, the LDOS inside the N--part is much
larger than that at the minima, and is very strongly modulated in the 
$z$-direction.  This is because constructive interference of the multiply
backscattered waves leads to a quasi-bound standing-wave state at the
conductance maxima, while the conductance minima are associated with a
condition of destructive interference
[compare Fig.\ \ref{fig.4}(b) (conductance minimum) and
Fig.\ \ref{fig.4}(c) (conductance maximum)]. For the first conductance
maximum after the channel opens as one widens the narrow section,
there is one maximum in the LDOS of the N--part in the longitudinal
direction; for the $n$th conductance maximum of the resonant
structure, there are $n$ longitudinal maxima of the LDOS. 
We thus see 5 maxima in Fig.\ \ref{fig.4}(c). 

The electronic structure of the nanowire shown in 
Fig.\ \ref{fig.4} gives vivid pictorial support to the claim
advanced in Ref.\ \onlinecite{sbb} that conductance channels should be
interpreted as mesoscopic bonds, which provide the cohesion of the system.
The claim advanced here that the electronic structure in such a nanowire
is dominated by quantum-confinement effects rather than by atomistic effects
is in agreement with STM studies of electron ``corrals'' on Cu surfaces.
\cite{corral}

\section{3D nanowire}
\label{sec.3dnanowire}

While the results for 2D nanowires presented in the preceding section are 
interesting both in illustrating the generality of the mesoscopic phenomena
in question and for their relevance to experiments on point contacts 
in quasi-2D electron gases,
the only experiments to date on the mechanical properties of nanowires 
\cite{nanoforce,nanoforce2} involve 3D metals.
In this section, we consider a 3D wire with WNW--geometry and square
cross--section.
For a square cross section, many modes
are doubly degenerate, as in the case of cylindrical symmetry, leading to
conductance steps of both $2e^2/h$ and $4e^2/h$. 
It would be possible to lift this
degeneracy by considering a wire with a rectangular rather than  square
cross--section.
The formalism to compute the
S--matrix (cf.\ Sec.\ \ref{sec.comp_smatr}) and to obtain the force,
charge, and conductance
(see Sec.\ \ref{sec.smatrix}) is the same as in the 2D
case, although one needs to include more evanescent 
modes for an accurate computation of the S--matrix than in the 2D case. 
It is straightforward to extend the present calculation
to wires of arbitrary cross--section, if the eigenfunctions and eigenvalues
of the 2D Schr\"odinger equation are known for that cross--section. 

Although the exactly solvable geometry considered here is somewhat special
due to the presence of sharp edges, 
the gross behavior of the conductance
and force is similar to that observed experimetally in 3D metallic nanowires
\cite{nanoforce,nanoforce2} and calculated
for smooth, adiabatic geometries.\cite{sbb}
The agreement of the present results for the WNW
geometry with the experimental results of Ref.\ \onlinecite{nanoforce}
is poorer than for the smooth geometries considered previously,
\cite{sbb} indicating that the experimental geometry is undoubtedly
much smoother than that considered here.  

\subsection{Force and Conductance}

\begin{figure}
\epsfxsize=14cm
\epsfysize=10.5cm
\hfil
\epsffile{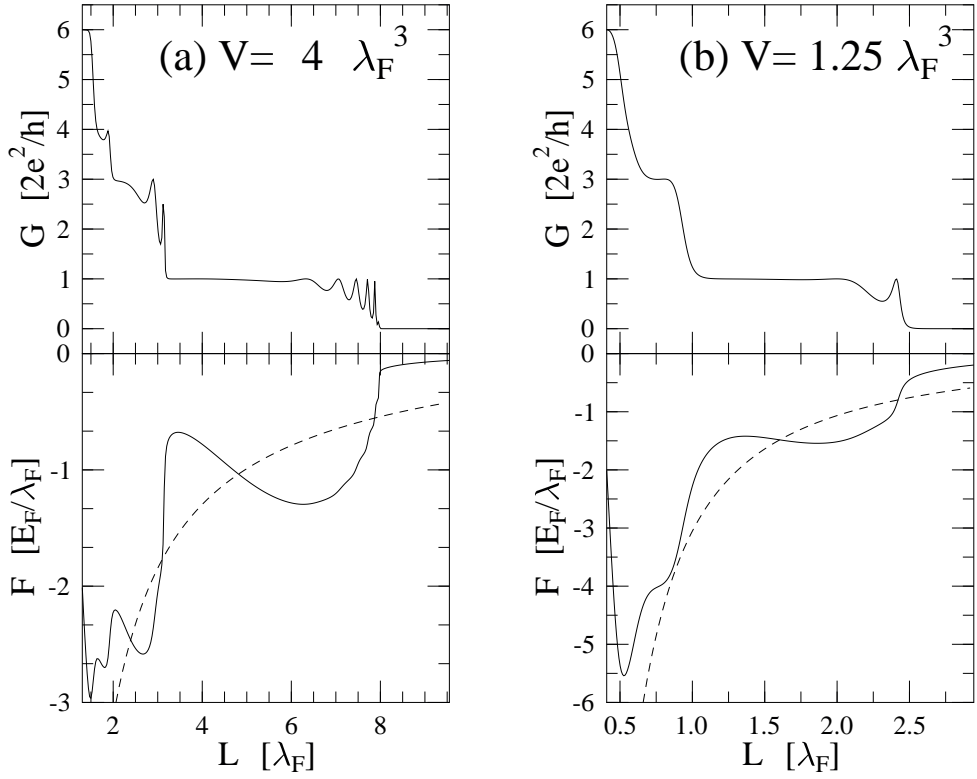}
\hfil
\caption{Electrical conductance and tensile force for two different 3D
wires with WNW--geometry and square cross--section as a function of
the length $L$ of the narrow part.  The width of the narrow part is determined 
by a constant--volume constraint $Ld^2=L_0D^2=V=\mbox{const}$, where $L_0$
is the initial length of the constriction before deformation.
In (a) the volume of the
narrow part is given by $V=4 \lambda_F^3$ while it is smaller ($V=1.25
\lambda_F^3$) in (b). The dimensions in (b) are comparable with those 
of the Au wire studied experimentally in Ref.\ 4.
The dashed curves give the force expected by the surface term plus
topological correction. }
\label{fig.5}
\end{figure}

Fig.\ \ref{fig.5} shows the conductance and tensile force as a function of 
elongation for two model 3D wires. 
Under elongation, the narrow section is assumed to deform plastically, {\it
i.e.,} its volume $V=Ld^2=L_0D^2$ is held constant, where $L_0$ is the initial
length of the narrow section.  The inclusion of an additional, small 
elastic deformation can be shown not to modify the mesoscopic effects in
an essential way.\cite{unpub}
The comparison of the exact results
shown here to the results of 
an adiabatic approximation is similar to that in the 2D case (c.f.\ Fig.\ 
\ref{fig.2}), so for clarity we have not shown them for the 3D case.

In Fig.\ \ref{fig.5}(a), a nanowire with a volume of $4\lambda_F^3$ is
shown, while a shorter wire with a volume of $1.25\lambda_F^3$ is shown in 
Fig.\ \ref{fig.5}(b).  The width of the wide sections is $D=1.76\lambda_F$,
which fixes the number of asymptotic propagating modes to be 6.
In Fig.\ \ref{fig.5}(a), one sees conductance plateaus at $G=1,3,4,6\times
G_0$, with a pronounced resonant structure superimposed due to multiple
reflection at the abrupt junctions between wide and narrow sections.  The 
sequence of degeneracies corresponds to the square symmetry of the 
cross--section  (cylindrical symmetry, on the other hand, gives
$G=1,3,5,6,\cdots\times G_0$, see Ref.\ \onlinecite{sbb}). 
Just as in the 2D case discussed above, the force exhibits mesoscopic 
oscillations of order $E_F/\lambda_F$, which are correlated with the 
conductance steps:  $|F|$ increases along the conductance plateaus, and drops
abruptly at the conductance steps.  The resonant structure in the conductance
is also reflected in the first derivative of the force, particularly on the 
last conductance plateau.  
The force is similar in magnitude to that calculated for a smooth geometry
in Ref.\ \onlinecite{sbb} and observed experimentally in Au nanowires in Ref.\ 
\onlinecite{nanoforce}: the forces required to cut off the last two conductance 
eigenmodes are of order $1.25E_F/\lambda_F$ and $2.5E_F/\lambda_F$,
respectively (recall that $E_F/\lambda_F \approx 1.7\mbox{nN}$ in Au).

The surface tension 
has been plotted for comparison as a dashed curve in Fig.\
\ref{fig.5}. It has been 
computed analogously to the 2D case from the Weyl expansion 
\cite{sbb,commentzwerger} 
of the grand canonical potential,
\begin{equation}
\Omega=-E_F\left(\frac{16\pi}{15\lambda_F^3} V 
-\frac{\pi}{4\lambda_F^2}\partial V + \frac{2}{3\lambda_F}\sum_{{\rm
edges\,} i} C_i L_i\right) + \delta \Omega,
\label{weyl.ohm3D}
\end{equation}
where $V$ is the volume and $\partial V$ the surface area of the nanowire;
the topological terms are proportional to the lengths of the edges
$L_i$, and the appropriate constants are $C_i = 1/4$ for an
edge with an inner angle of $\pi/2$ and $C_i=-5/36$ for an
edge with an inner angle of $3 \pi/2$.  The surface tension, or semiclassical
approximation to the force, is obtained from the derivative of the 
semiclassical approximation to $\Omega$ [the term in parentheses in Eq.\
(\ref{weyl.ohm3D})] with respect to $L$, which yields
\begin{equation}
F_{\rm surf}=-\frac{E_F}{\lambda_F}\left(\frac{\pi d}{2\lambda_F}
+\frac{\pi d^2}{2L\lambda_F} -\frac{2}{3} - \frac{d}{2.7 L}\right)
\label{fsurf.3D}
\end{equation}
for a 3D nanowire with WNW geometry and square cross--section.
Aside from the initial deformation, where the topological force
is important (c.f.\ Sec.\ \ref{sec.osc}), the force exhibits oscillations
centered about the semiclassical result (dashed curve).

In Fig.\ \ref{fig.5}(b), a shorter 
wire whose conductance versus elongation matches
the experimental curve shown in Fig.\ 1 of Ref.\ \onlinecite{nanoforce}
is shown; $L_0$ was chosen such that the elongation required to decrease the
conductance from $6 G_0$ to 0 is $2\lambda_F \approx 1\mbox{nm}$.
We see that the conductance and force are correlated in a similar way.
Due  to the shorter length of the constriction, the
conductance steps are smeared out by tunneling 
and by above-threshold reflection; the plateau at $G=4G_0$
and the associated structure in the force are no longer visible.
The resonant structure in the conductance is also suppressed, except on the
last plateau, where the narrow section is longest.
The overall magnitude of the force is larger
than for longer constrictions, due to the increased surface tension, and the 
total elongation required to break the nanowire is less.
(Note that the effective
surface tension can be reduced by up to a factor of 5
by including a small elastic deformation.\cite{unpub})
However, the oscillations of the force around the
semiclassical approximation are of the same order as in the previous
case (see Fig.\ \ref{fig.q3}).  
The pronounced difference in Figs.\ \ref{fig.5}(a) and \ref{fig.5}(b) 
indicates a breakdown of the invariance of $F$ under a stretching of the
geometry $d(z)\rightarrow d(\lambda z)$, which was found within the WKB
approximation,\cite{sbb} due to strong tunneling effects in very short wires.

As in the 2D case, the force decays to zero
with increasing elongation after the last conductance channel is cut
off, although more slowly than does the conductance itself (see Sec.\
\ref{force_and_cond} for a discussion). 
$F$ remains non-negligible even for the largest elongations shown in Fig.\ 
\ref{fig.5}, when the conductance is exponentially small.\cite{remark.jerome}
A similar effect was observed experimentally  
(c.f.\ Fig.\ 1 of Ref.\ \onlinecite{nanoforce}), although it is not clear
whether the effect was above the noise level.

\subsection{Charge oscillations and screening}
\label{sec.screen}

The charge on such a 3D nanowire may be calculated from Eq.\ (\ref{qfromS}),
as in the 2D case.
The charge on the nanowire
changes as the wire is elongated due to surface terms and mesoscopic 
oscillations.  The Weyl expansion for the electronic charge of the nanowire is
\begin{equation} 
\langle Q^-\rangle
=-e \left(\frac{ 8 \pi}{3 \lambda_F^3} V - \frac{ \pi}{2 \lambda_F^2} \partial V
+ \frac{1}{\lambda_F}
\sum_{{\rm edges\,} i} C_i L_i \right ) + \delta Q_0,
\label{weyl.q3}
\end{equation}
The term in parentheses in Eq.\ 
(\ref{weyl.q3}) varies smoothly as the geometry of the wire is altered, while
$\delta Q_0$ describes the mesoscopic oscillations associated with the opening
or closing of discrete transverse modes.

\begin{figure}
\epsfxsize=14cm
\epsfysize=10.5cm
\hfil
\epsffile{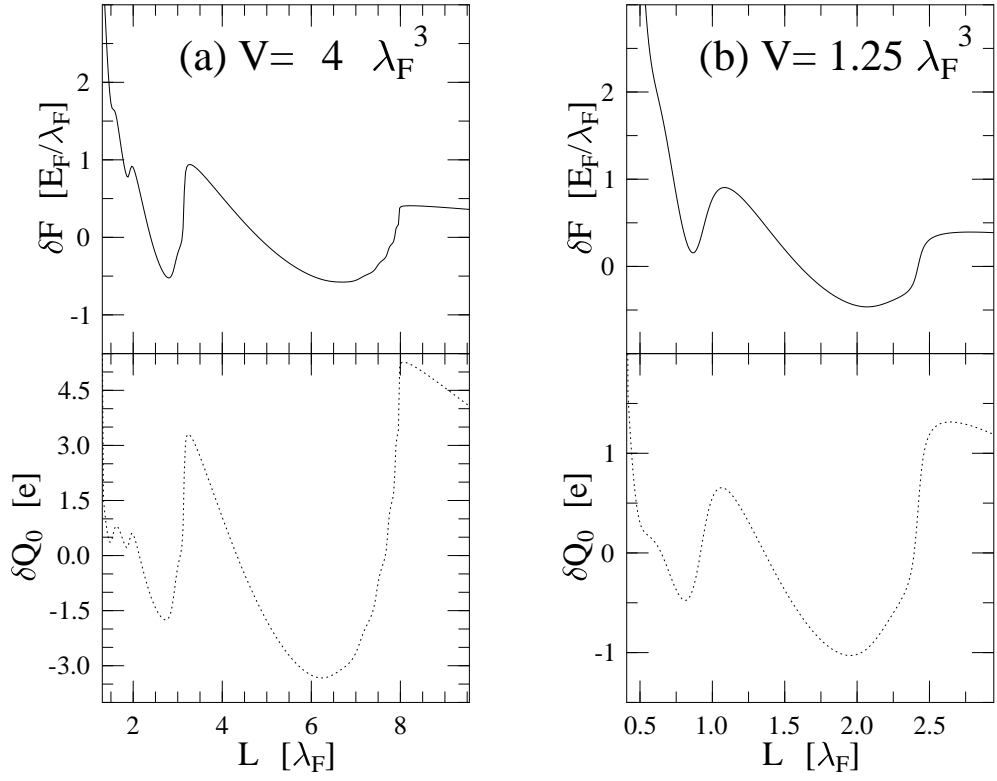}
\hfil
\caption{Charge and force oscillations as a function of elongation for
two 3D wires with WNW geometry. The same parameters as in
Fig.\ \ref{fig.5} have been used.}
\label{fig.q3}
\end{figure}

In Fig.\ \ref{fig.q3}, the mesoscopic charge oscillations  $\delta Q_0$,
calculated by subtracting the term in parentheses in Eq.\ (\ref{weyl.q3})
from the exact charge computed via Eq.\ (\ref{qfromS}),
and the force oscillations, calculated by subtracting the surface tension from
the total force, are plotted for both wires shown in Fig.\ 
\ref{fig.5} as a
function of elongation. As in the 2D case, there is a strong
correlation between the two quantities, and the charge oscillations are of
order the fundamental quantum of charge $e$. 
The force oscillations are, as in the case of an adiabatic geometry studied
in Ref.\ \onlinecite{sbb}, of order $E_F/\lambda_F$, aside from the 
nonuniversal topological correction occuring for small deformations from
an ideal wire, which was discussed in detail for the 2D case in Sec.\
\ref{sec.top}.

In an interacting system, screening of the charge oscillations will
occur. The net charge $\delta Q$ including screening
can be estimated within the Thomas-Fermi approximation as follows:
\begin{equation}
\label{deltaQ}
\delta Q = \delta Q_0- e^2 D(E_F) \delta V,
\end{equation}
where $\delta Q_0$ are the charge oscillations in the noninteracting
case regarded above, $D(E)$ is the density of states (integrated over
the length of the constriction) and the potential $\delta V$ 
due to the charge imbalance on
the wire can be estimated as
\begin{equation}
\label{qpotential}
\delta V=\frac{ \delta Q}{C}.
\end{equation}
We have introduced a phenomenological quantity $C$ corresponding to the
total
capacitance of the inner part of the wire to its surroundings.
Eqs.\ (\ref{deltaQ}) and (\ref{qpotential}) can be used to compute 
self-consistently the charge on the wire within linear response:
\begin{equation}
\delta Q = \frac{ \delta Q_0}{1+e^2 D(E_F) / C}.
\label{q.screened}
\end{equation}
The charge in the noninteracting case $\delta Q_0$ has already been computed
and discussed (see Fig.\ \ref{fig.q3}). 

\begin{figure}
\epsfxsize=14cm
\epsfysize=10.5cm
\hfil
\epsffile{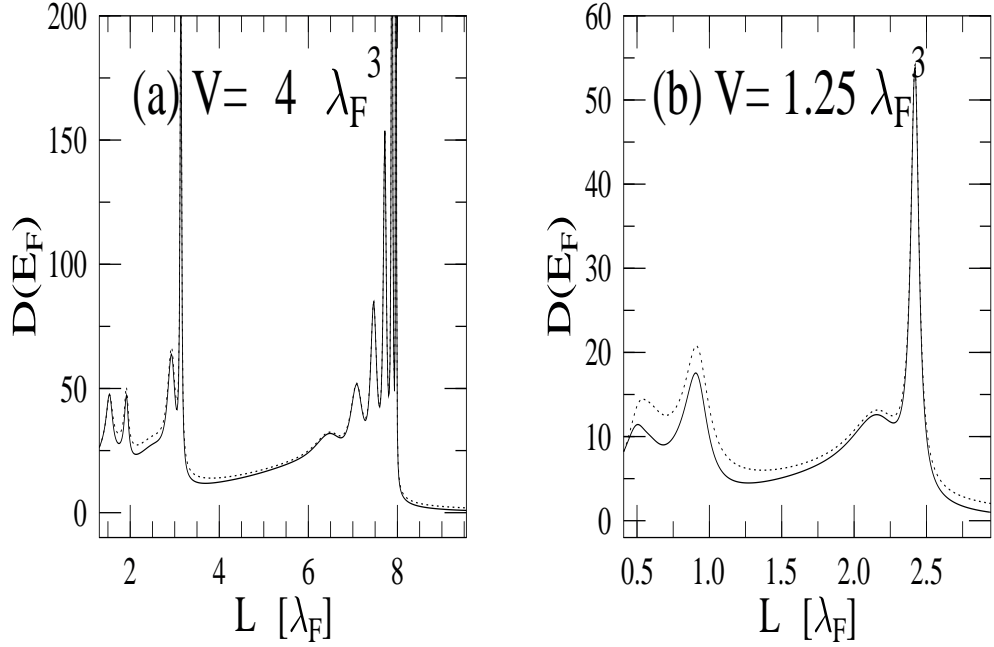}
\hfil
\caption{Density of states at the Fermi surface as a
function of elongation for 3D wires with parameters as in
Fig.\ \ref{fig.5}. The solid line shows the LDOS integrated over the
narrow part of the wire, the dotted line the DOS 
computed from Eq.\ (\ref{DOS}).}
\label{fig.q4}
\end{figure}

On dimensional grounds, the capacitance obeys $C=\alpha L$, where $\alpha$
is a geometrical constant of order 1 ($\alpha$ may depend logarithmically
on the ratio $D/L$).
The density of states can be computed by a spatial integral
of the LDOS over the narrow section, or from the asymptotic scattering
phase shifts via Eq.\ (\ref{DOS}).  The later definition includes the 
contribution of the Friedel oscillations induced in the wide sections.
In Fig.\ \ref{fig.q4}, both densities of states
are shown for wires with the same parameters as above. They are
approximately equal, indicating that the excess charge $\delta Q$ induced on
the nanowire under deformation resides mainly on the narrow section.
The intricate resonant structure in $D(E_F)$ occurs due to the formation of
quasi-bound states due to multiple reflection at the junctions of the wide
and narrow sections (see Fig.\ \ref{fig.4}), and would not be present for
a smooth geometry, such as that studied in Ref.\ \onlinecite{sbb}.
Aside from this resonant structure,
the overall magnitude of $D(E_F)$ can also be determined from a
Weyl expansion.  To leading order, one finds $D(E_F) \sim (4L/E_F\lambda_F)
G/G_0$.  Inserting this expression into Eq.\ (\ref{q.screened}),
one finds
\begin{equation}
\delta Q \sim \frac{\delta Q_0}{1+0.66 r_s G/G_0\alpha},
\label{q.screened.3D}
\end{equation}
were $r_s$ is the dimensionless electron gas parameter, which takes values 
between 2 and 6 in metals.  This indicates that the screening
of the mesoscopic charge fluctuations is poor for wires with small
conductance.  The screened mesoscopic charge fluctuations should be 
measurable with a local probe, such as a single-electron transistor.

The screening of the predicted charge fluctuations should be even weaker
in 2D GaAs quantum point contacts, due to the large dielectric constant,
which enhances the capacitance.  In the 2D case, the constant 0.66 is
replaced by 0.52 in the denominator of Eq.\ (\ref{q.screened.3D}), and
the geometrical factor $\alpha\rightarrow \varepsilon \alpha$, where
$\varepsilon\approx 13$.  This indicates that the predicted charge 
fluctuations are essentially unscreened for small-conductance QPCs in
GaAs.

Let us finally add a comment on the effect of screened electron-electron
interactions on the free energy.  Within linear response, the Coulomb
energy associated with the mesoscopic charge imbalance is given by
\begin{equation}
\Delta \Omega_C = \frac{1}{2} \int d^3 x \, \vec{D}\cdot\vec{E}.
\end{equation}
In the discrete potential model introduced above to treat screening,
this leads to
\begin{equation}
\Delta\Omega_C = \frac{\delta Q_0^2}{C+e^2 D(E_F)}.
\label{omega.C}
\end{equation}
The details of the derivation of Eq.\ (\ref{omega.C}) will be given elsewhere.
\cite{unpub} Equation (\ref{omega.C})
indicates that the mesoscopic charge fluctuations of order $e$ lead to
a negligible correction to the free energy of the system, even in the limit 
$C\rightarrow 0$ of perfect screening, justifying the independent-electron 
model of nanocohesion.
 
\section{Conclusions}
\label{sec.concl}

In the present article, we have investigated the conducting and thermodynamic
(including mechanical) properties of metallic nanowires with a 
wide--narrow--wide geometry, using a free-electron model with hard-wall
boundary conditions.  All properties of the nanowire were related to the 
electronic scattering matrix, which was evaluated exactly, including all
effects of tunneling and interchannel scattering.  The present results 
confirm the central conclusion of Ref.\ \onlinecite{sbb}, which was based
on an evaluation of the scattering matrix within the adiabatic and WKB
approximations, that closing a conductance channel by stretching a metallic
nanowire requires a force of order $E_F/\lambda_F$, or roughly
a nanonewton in monovalent metals, independent of the total number of 
conducting channels.  

In contrast to this ``universal''
behavior under a smooth deformation of the geometry, 
we have shown that the force associated with a change in topology
can be large compared to $E_F/\lambda_F$, and
indeed comparable to the total (macroscopic) cohesive force.

In addition, we predict that the net charge
on a nanowire exhibits oscillations of order the fundamental charge quantum
$e$, which are synchronized with the force oscillations and conductance
steps.  These charge oscillations should also be present in quasi-2D 
quantum point contacts in GaAs heterostructures, and should be experimentally
detectable using a local probe, such as a single-electron transistor.

A final word should be added by way of addressing the central controversy
of this field, which can be stated as follows:  
Is it the atomic structure ({\it i.e.}, the bonds) which determines the 
conductance,\cite{landman,todorov,mol.dyn,nature} or should the conductance
channels themselves be thought of as mesoscopic bonds which provide the 
cohesion, and thus determine the structure?  

The conductance channels are the eigenstates of the electronic scattering
problem,\cite{eigenchannels} and are thus the appropriate states to describe
both the transport and thermodynamic properties of a nanowire, which is 
an open quantum mechanical system.  These
scattering states are linear combinations of local bonding states, so there
is no fundamental contradiction between the two viewpoints stated above:
one is free to look at the problem in a localized basis of bonds
or in a basis of extended electronic eigenstates.  

However, since an exact
solution of the many-body Schr\"odinger equation for a nanowire is 
beyond our current capabilities, one is forced to make certain approximations,
which are convenient in the basis of choice.  Thus molecular dynamics
simulations\cite{landman,todorov,mol.dyn} typically neglect any quantum
coherence between different bonds, and amount to a computational version of
the classical
ball--and--stick model of atomic structure, where bonds are described by
a short-ranged empirical inter-atomic potential.  This is an uncontrolled
approximation, which should be adequate to describe covalent bonds in an
insulator, but its applicability for monovalent metals
with nearly spherical Fermi surfaces like Au and Na is questionable.

The molecular dynamics simulations involve
empirically determined short-range interatomic potentials whose characteristic
length and energy scales mimic the quantum mechanics of bonding.  When
playing classical mechanics with these quantum forces, it is not too 
surprising if one obtains forces of the right order of magnitude.  
Such models are of course inadequate 
to describe electrical conduction, so to explain the observed correlations
in the conductance and force of metallic nanowires, a quantum-mechanical 
model whose geometry is fit to the results of the classical simulation is
constructed.\cite{landman,todorov,mol.dyn}
The cost of working in a localized basis is thus the necessity of using
{\em different physical laws} to describe conductance and cohesion.

On the other hand, we have seen in the present article (see also Ref.\
\onlinecite{sbb}) that the observed correlations in the 
conducting and mechanical properties of metallic nanowires can be accounted
for naturally in a {\em single} quantum mechanical model, which treats the 
mechanical and electrical properties of the system on an equal footing.  In 
order to solve the quantum scattering problem, we have neglected the 
discrete atomic structure, working in a jellium-like model, which is equivalent
to assuming that the only effect of the lattice is to modify the electron's
effective mass.  
This should be a rather good approximation for simple metals
like Na and adequate for noble metals like Au.
Defects in the atomic structure of the wire or roughness
in its surface introduce additional scattering, which can also be included in
the jellium model in a natural way.\cite{jerome}
A drawback of the jellium model, or at least of the assumption employed
here and in Ref.\ \onlinecite{sbb} that the positive background deforms
continuously as the nanowire is elongated, is the inability to describe the 
hysteretic behavior found in the experiment of Ref.\ \onlinecite{nanoforce}.
The claim advanced here that the electronic structure in such a nanowire
is dominated by quantum-confinement effects rather than by atomistic effects
is in agreement with STM studies of electron ``corrals'' on Cu surfaces.
\cite{corral}

In the end, the merits of the jellium model vis \`a vis an atomistic 
description must be decided based on its predictive power.  An interesting
prediction of the jellium model discussed above is the existence of
mesoscopic charge fluctuations of order $e$, which are strongly correlated
with the force fluctuations, and synchronized with the conductance steps.
These charge fluctuations are a collective effect, which can not be
described in a ball--and--stick picture of bonding.  It is incumbent on the 
experimenter to verify or falsify this clear prediction of the jellium
model.

\begin{center}
{\bf ACKNOWLEDGEMENTS}
\end{center}

C.\ A.\ S.\ acknowledges valuable discussions with J.\ B\"urki and D.\
Baeriswyl.  This work was supported in part by the Deutsche 
Forschungsgemeinschaft through grant SFB 276.

\pagebreak

\pagebreak

\pagebreak

\begin{references}
\bibitem{remark.jerome} We wish to thank J. B\"urki for useful comments on
this point.
\end{references}
\end{document}